Photonic stop bands in quasi-random nanoporous anodic alumina structures

Ivan Maksymov[1], Josep Ferré-Borrull[2], Josep Pallarès[2], Lluis F. Marsal[2]

1.- Nonlinear Physics Center, Research School of Physics and Engineering, Australian National University, Canberra ACT 0200, Australia.

2.- Universitat Rovira i Virgili, Nano-electronic and Photonic Systems, Av. Paisos Catalans 26, 43007, Tarragona, Spain, Fax: +34977559605, josep.ferre@urv.cat.

## Abstract

The existence of photonic stop bands in the self-assembled arrangement of pores in porous anodic alumina structures is investigated by means of rigorous 2D finite-difference time-domain calculations. Self-assembled porous anodic alumina shows a random distribution of domains, each of them with a very definite triangular pattern, constituting a quasi-random structure. The observed stop bands are similar to those of photonic quasicrystals or random structures. As the pores of nanoporous anodic alumina can be infiltrated with noble metals, nonlinear or active media, it makes this material very attractive and cost-effective for applications including inhibition of spontaneous emission, random lasing, LEDs and biosensors.

## 1. Introduction

The proposal of photonic crystals (PhCs) by Yablonovitch and John[1,2] have extended the research into the field of nanophotonics devices benefiting from the formation of a photonic band-gap (PBG) and the inhibition of the spontaneous emission of light. These remarkable physical properties have made PhCs very promising for a wealth of applications such as lasers[3], waveguides[4], photonic circuits[5] or optical cloaking[6]. Further investigation has been motivated by the search for other materials with similar physical properties, but that would offer advantages over PhCs for many applications even when the refractive index contrast among the constituent materials is low. One of the solutions has been found in the use of one[7] and two-dimensional[8] photonic quasicrystals (PQCs). PQCs have a higher degree of rotational and point-reflection symmetry than conventional PhCs. These symmetries facilitate the formation of the PBGs even in materials with low refractive index contrasts. These PBGs are significantly more isotropic and bands of their dispersion relations are flat[9], which is advantageous for use as highly efficient isotropic thermal radiation sources[10]. PQCs structures can support a rich variety of localized modes (such as constant flux modes[11]) as well as take advantage of Anderson localization phenomena[12]. Consequently PQCs can be employed to design random lasers with tailored field patterns with possible application to biological sensing [13]. In LED applications, PQCs help to extract light and boost LED emission characteristics [14].

In this work we pay attention to the self-assembled nanoporous anodic alumina[15] (NAA), a material thoroughly investigated and widely applied in nanotechnology[16]. To our best knowledge, no paper has been published to study the PBG properties of an NAA structure. NAA is obtained by the electrochemical etching of aluminum.

Under proper conditions its nanoporous structure shows a self-ordered lattice with a two-dimensional triangular periodic arrangement[15]. The lattice constant can be tuned from some tens of nanometers up to some hundreds. Although this triangular arrangement can be achieved over a long range by nanoimprinting the aluminum[17], in self-assembled NAA the lattice is broken into domains with a size of some tens of lattice sites and randomly oriented. One can see this as a quasi-random pattern, similar to a polycrystal, random in nature but with some degree of order. Figure 1 shows an example of an NAA structure obtained with the conditions described in Ref. 18.

Hereafter we investigate the formation of photonic stop-bands in quasi-random NAA structures. A superposition of the stop-bands from the triangular periodic arrangement of their domains leads to the formation of a general PBG. Given that the orientation of the domains is random, the investigated structures are isotropic in a macroscopic sense. This isotropy should favor the formation of a PBG in the same way the higher rotational symmetry of photonic quasicrystals produce PBGs for smaller index contrasts than for periodic structures. However, the higher rotational symmetry of NAA structures is counteracted by the finite size of the domains, and thus, it is necessary to investigate to what extent this is affecting the gap formation.

## 2. Computational Approach

In our study of the stop-band formation in NAA structures we employ a 2D finite-difference time-domain (FDTD) method. The FDTD is a grid-based numerical method for simulating the propagation of electromagnetic waves in arbitrary media [19]. This makes the FDTD especially attractive for calculating transmission properties of NAA structures. A triangular pore arrangement of the NAA domains facilitates the analysis because their photonic properties are in close physical analogy to those of 2D perfect triangular-lattice photonic crystals, whose physics is well understood to a certain extent. Hereafter we will perform systematic 2D FDTD calculations assuming that the investigated structure is infinite in the direction normal to the apertures of pores in anodic alumina. We will restrict ourselves to the *H*-polarization, for which the magnetic field is oriented along the *z*-axis (Fig. 2).

Despite a certain analogy with perfect periodic structures, we cannot employ computational approaches widely used in the PhC research relying on the use of periodic and symmetric boundary conditions simplifying the calculation. In order to study the photonic properties of NAA structures, we use the structural data from experimentally obtained samples such as that of Fig.1 and perform statistical calculations over a set of randomly chosen domains. The procedure is as follows: the pore positions are determined by processing a SEM picture of a NAA sample with mathematical morphology algorithms. The average interpore distance *a* will be used as unit length in what follows. Then, four sets each one consisting of 20 subdomains of *13a×13a*, *18a×18a*, *22a×22a* and *26a×26a* respectively, are randomly chosen from these data and ideal circular pores (hereafter referred to as scatterers) with radius of *r=0.30a* are placed at the corresponding pore positions. FDTD calculations enable evaluation of the transmittance of each subdomain. An example of a *18a×18a* subdomain is schematically presented in Fig. 2 showing the positions of the incident plane-wave source, the NAA structure and the detectors lined up behind it. In Fig. 2 gray circles denote the scatterers of the NAA structure (refractive index *n*=1) whereas the white background denotes the host material (alumina, *n*=1.65). Convolution Perfectly Matched Layers (CPML) absorbing boundary conditions[19] are applied to compute the fields in unbounded regions of the model. To evaluate the transmittance of a given subdomain, the spectrum of the transmitted wave is normalized to the spectrum of the same wave transmitted through the host medium at the same distance.

The finite-difference grid used in our calculations is composed of square cells with edge size of $\Delta=0.042a$. Proper care is taken to ensure that the time step $\Delta t$ satisfies the Courant stability criterion [19] so that $\Delta t=\Delta/(2c)$, where $c$ is the light speed in vacuum. A subgrid is used to exactly specify the size of the scatterers and smooth their contours that suffer from a staircase effect inherent in calculating with a spatial resolution of 24x24 grid cells per scatterer used in our model. We use 20 CPML layers in each coordinate direction with a polynomial grading for the conductivity suggested in Ref. 18. The initial condition is posed so that a plane wave is excited in front of the investigated NAA structure (Fig. 2). The plane wave is formed by a line of in-phase pointwise 'hard' sources that provide a bandpass Gaussian pulse with zero dc content. The spectrum of the source covers the frequencies between $\omega a/2\pi c=0.1$ and $\omega a/2\pi c=0.8$ with enough intensity. The computation time is long enough (50 000 iterations) to guarantee a decay of the fields inside the computational domain, so that the transmitted spectra do not change appreciably for longer computational times. To obtain a transmitted spectrum, the Poynting vector flux is calculated from data obtained by means of a Fast Fourier transformation (FFT) algorithm.

In order to deeper understand the formations of the stop-band in the investigated NAA structure, we calculate the average of the transmittance for all the subdomains considered in the calculations, $\bar{t}_i(\omega)=1/n\sum_{j=1}^{n} t_{i,j}(\omega)$, where $i = 1,\ldots,4$ denotes the subdomain size as defined hereinabove, $j$ denotes one subdomain of a given size and $n=20$. The standard deviation of the transmittance for a given subdomain size, defined as $Std_i(\omega) = \left(1/n \sum_{j=1}^{n} [t_{i,j}(\omega) - \bar{t}_i(\omega)]^2\right)^{1/2}$ [20], is also evaluated.

## 3. Simulation Results

Fig. 3 shows the four calculated average transmittances. The ranges corresponding to the PBGs for a triangular lattice of air cylinders in alumina, with the same radius and for both high symmetry directions, are indicated. The graphs show the existence of a low-transmittance region in the spectra (a stop-band) in the same range an equivalent periodic structure with the lattice constant *a* shows a PBG. The inset (a) in Fig. 3 shows the same $\bar{t}_i(\omega)$ in the stop band region. It can be observed that for increasing subdomain size, the width of the stop band increases while the minimum transmittance decreases. The inset (b) in Fig. 3 shows the transmittance spectra standard deviation. These standard deviations show a low value in the same range as the low average transmittances. This can be interpreted as that the variability of the transmittance among different subdomains is smaller in the stop-band region.

## 4. Conclusions

The results obtained with accurate 2D FDTD calculations convincingly support our assumption that the arrangements of self-assembled pores in porous anodic alumina structures demonstrate pronounced stop band properties similar to those of photonic quasicrystals or random structures. Taking into account that the self-assembled nanoporous anodic alumina can be obtained without any costly prepatterning process such as lithography or nanoimprinting, and that the structure can be easily transferred to other materials such as silicon or silicon dioxide, or its pores can be filled with polymers, polymer-nanoparticle composites, noble metals, optical gain materials or nonlinear material, we anticipate the use of this promising material becoming more widespread for the inhibition of spontaneous emission, random lasing and spasers[12] among others.

**Acknowledgements**

The authors gratefully acknowledge the support of Spanish Ministry of Ciencia e Innovacion (MICINN) under grant number TEC2009-09551, HOPE CSD2007-00007 (Consolider-Ingenio 2010) and AECID-A/024560/09. I.M. IM acknowledges the support of the Australian Research Council.

**Figure Captions**

Figure 1:

SEM Picture of a NAA structure used to extract the pore positions considered in the simulations.

Figure 2:

(Color online) Schematic of the computational domain. Gray circles denote the scatterers of the investigated NAA subdomain, white background corresponds to the alumina.

Figure 3:

Average transmittance obtained from the transmittance of 20 subdomains for different sizes of NAA structure considered. Top inset: a detail of the range with the smallest transmittance values. Bottom inset: standard deviation of the transmittance for the 20 subdomains.

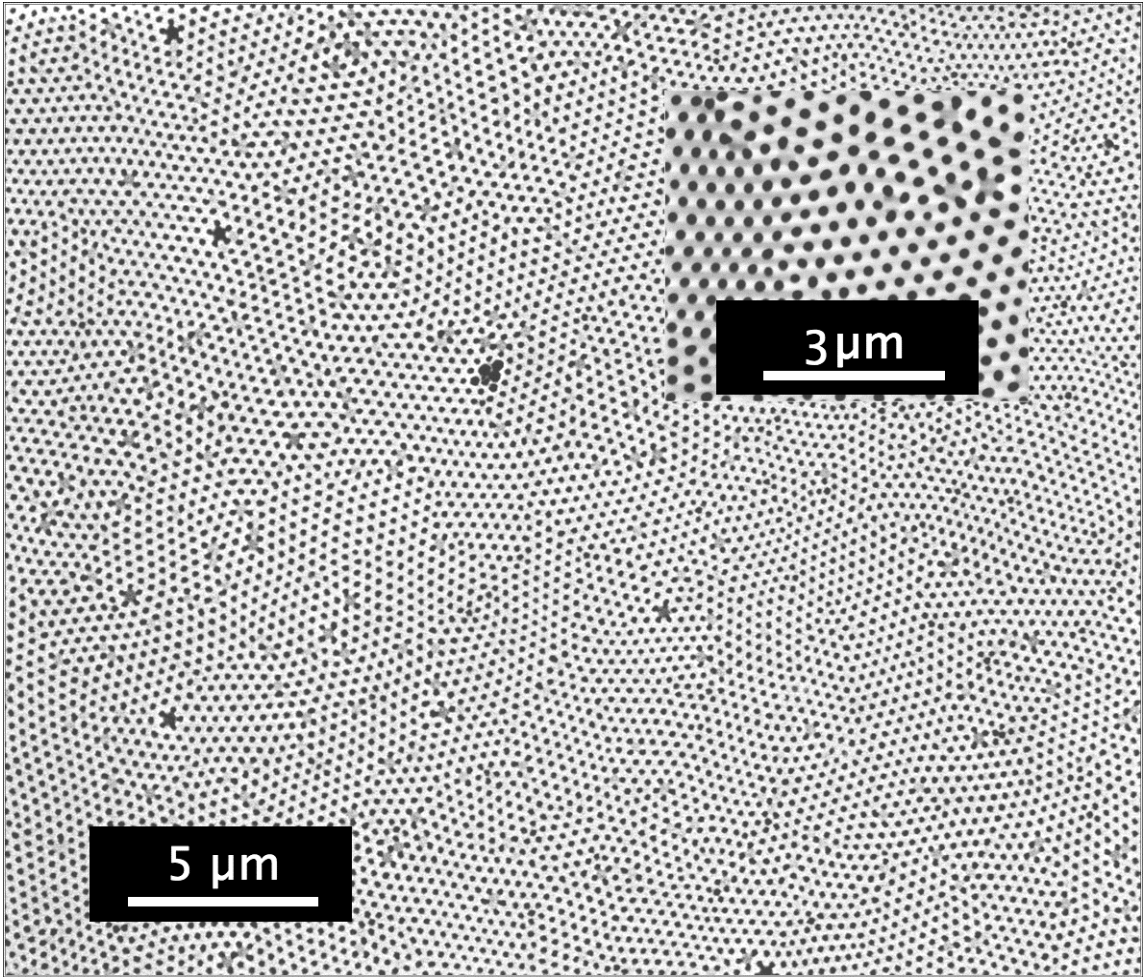

Figure 1

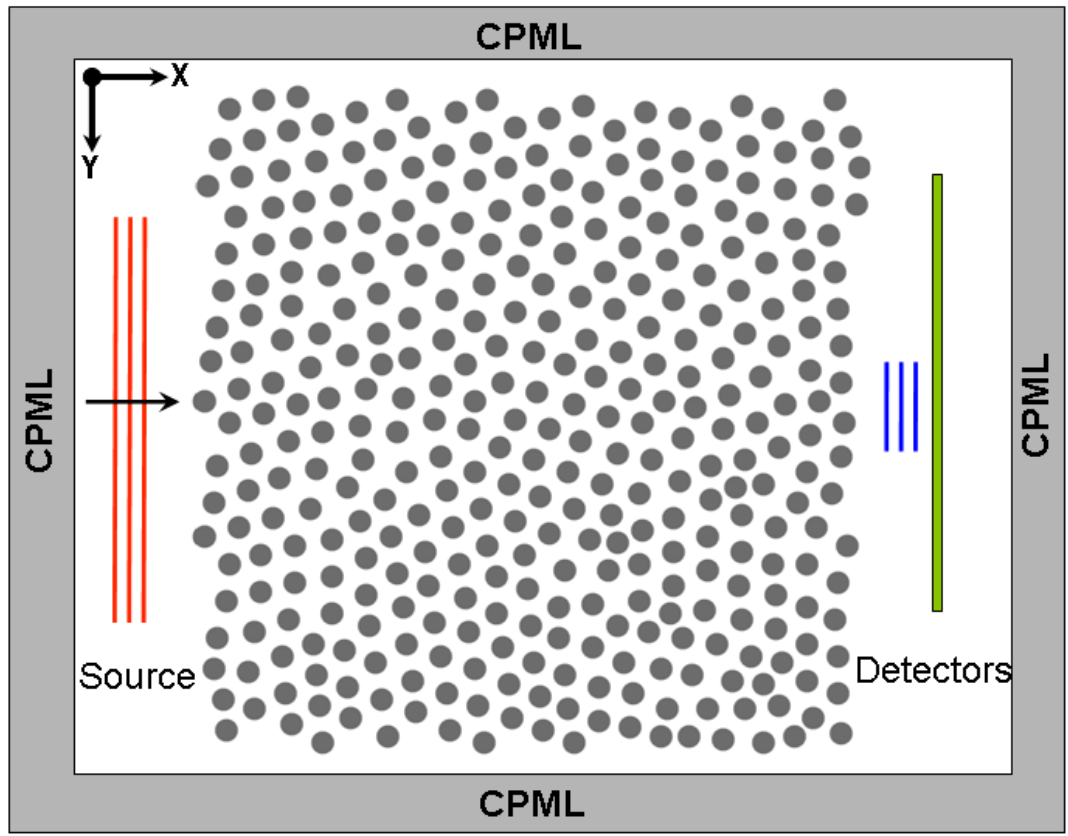

Figure 2

**Figure 3**

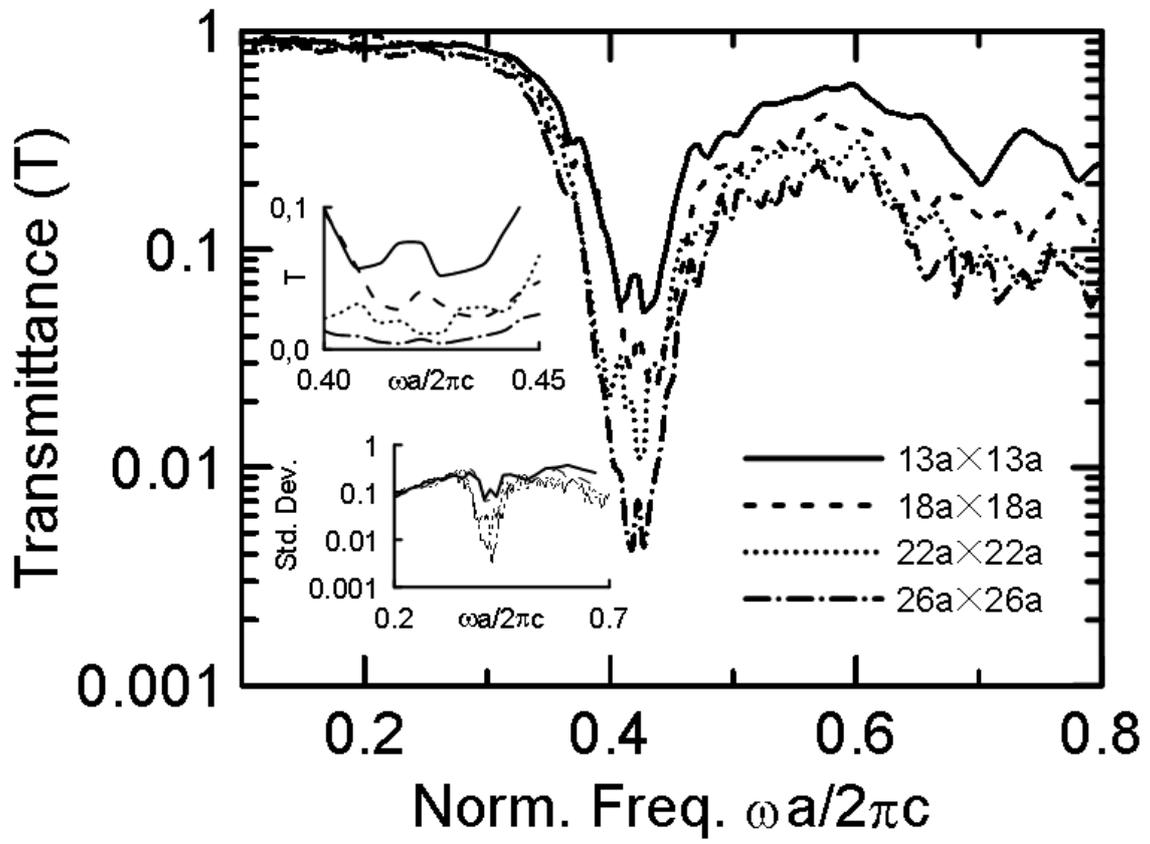

Figure 3